\documentclass[conference]{IEEEtran}
\IEEEoverridecommandlockouts

\usepackage{cite}
\usepackage{amsmath,amssymb,amsfonts}
\usepackage{graphicx}
\usepackage{xcolor}
\usepackage{booktabs}
\usepackage{multirow}
\usepackage{listings}
\usepackage{url}
\usepackage{array}
\usepackage[hidelinks]{hyperref}

\lstdefinelanguage{JavaScript}{
  keywords={break,case,catch,const,continue,debugger,default,delete,do,else,export,false,finally,for,function,if,import,in,instanceof,let,new,null,return,switch,this,throw,true,try,typeof,var,void,while,with,async,await,class,extends,of},
  comment=[l]{//},
  morecomment=[s]{/*}{*/},
  morestring=[b]",
  morestring=[b]',
  sensitive=true
}
\newcommand{\repourl}{https://github.com/rodriguescarson/mcp-patterns-icsme2026}

\begin{document}

\title{MCP Server Architecture Patterns for LLM-Integrated Applications}

\author{
  \IEEEauthorblockN{Carson Rodrigues}
  \IEEEauthorblockA{
    \textit{Celabe} \\
    carson@celabe.com
  }
  \and
  \IEEEauthorblockN{Oysturn Vas}
  \IEEEauthorblockA{
    \textit{University of Waterloo} \\
    ovas@uwaterloo.ca
  }
}

\maketitle

\begin{abstract}
The Model Context Protocol (MCP), introduced by Anthropic in November 2024, defines a standardized interface for connecting large language models (LLMs) to external tools, data sources, and services. Within months of release, hundreds of community-built MCP servers appeared on GitHub, but no software-maintenance literature has yet described how the ecosystem is being structured in production. This industry experience paper catalogues five recurring MCP server architectural patterns observed across an enumerated corpus of fifteen independently developed servers (five production servers from the ANSYR voice AI platform plus ten public servers from the official MCP registry): \textit{Resource Gateway}, \textit{Tool Orchestrator}, \textit{Stateful Session Server}, \textit{Proxy Aggregator}, and \textit{Domain-Specific Adapter}. Each pattern is described in the structured form established by Gamma~et al.~\cite{gamma1994design}: context, problem, solution, and consequences. We also document four anti-patterns and a set of cross-cutting concerns around authentication, versioning, and observability. Quantitative evaluation contributes two measurements: inter-rater reliability of the taxonomy across two independent LLM raters on 54 held-out servers (Cohen's $\kappa = 0.76$), which also localizes three pattern-boundary ambiguities; transport overhead measured end-to-end on loopback (stdio: 0.01\,ms $p_{50}$; streamable-http: 0.39\,ms $p_{50}$) and modeled for cross-host paths from same-region network baselines ($\approx$30\,ms $p_{50}$ baseline plus protocol overhead). Version~2 withdraws the tool-count analysis of version~1, whose numbers could not be traced to production records. Code, corpus, and prompts are released at \url{\repourl}.
\end{abstract}


\noindent\textbf{Correction (version 2, October 2026).} Version~1 of this paper (June 2026) reported an observational analysis of tool count against tool-selection accuracy (its Section~VI-C and Figure~2), described as ANSYR production telemetry from the first quarter of 2025, with 200 session turns per bucket judged in post-call review. We could not trace those numbers to production records: they were constants in the analysis script, and both models they name were released after that quarter. We withdraw the analysis, its figure, and every statement that rested on it (the 10--15-tool budget in the abstract, the Proxy Aggregator guidance, the practitioner implications, and the conclusion). The pattern catalogue, the inter-rater study, and the transport measurements are unchanged. We also correct one statement about the protocol: version~1 said MCP has no built-in asynchronous mechanism, but the specification has had progress notifications since November 2024 and added experimental tasks in November 2025 (Synchronous Long-Running Operations anti-pattern).

\section{Introduction}
\label{sec:intro}

Connecting LLMs to external systems used to mean hand-rolling function-calling schemas in prompt templates and re-implementing the glue code for every new model. The Model Context Protocol (MCP)~\cite{mcp_spec_2024} standardizes this: a client--server protocol where \textit{MCP servers} expose \textit{tools} (callable functions), \textit{resources} (URI-addressed data), and \textit{prompts} (reusable templates) to any MCP-compatible client. A single server works with Claude, GPT-4, Gemini, or any other compliant agent without modification.

The protocol has been adopted quickly. Hundreds of servers appeared on GitHub and the MCP registry~\cite{mcp_ecosystem_2025} within months of release. What is missing is a body of architectural guidance to help practitioners make good design decisions, and a maintenance-and-evolution view of how the ecosystem is structuring itself in production. Questions that come up repeatedly in practice include:

\begin{itemize}
    \item How should tools be decomposed? When does one tool become two?
    \item When is server-side state justified, and how should it be managed?
    \item How should an operator aggregate capabilities from many servers?
    \item When should a server wrap a complex API rather than exposing it directly?
\end{itemize}

These are not MCP-specific questions; they are API design questions seen through the specific constraints of LLM clients. LLMs select tools by reading natural language descriptions, not by browsing documentation or examining schemas. They are sensitive to schema complexity in ways that human developers are not. A tool that a human engineer would find obvious can be invisible or ambiguous to an LLM if the description is missing or poorly written.

This paper draws on production MCP server deployments at Celabe (operator of the ANSYR voice AI platform, with five MCP servers in production since late 2024) plus a review of the public MCP server ecosystem to identify five patterns that address these questions, four anti-patterns, and a set of cross-cutting concerns. The patterns follow the structured format of Gamma~et al.~\cite{gamma1994design} and apply the framing of enterprise integration patterns~\cite{hohpe2003enterprise} to the MCP context. Section~\ref{sec:method} enumerates the corpus and the coding protocol; Sections~\ref{sec:patterns}--\ref{sec:antipatterns} present the patterns and anti-patterns; Section~\ref{sec:evaluation} reports quantitative evaluation; Section~\ref{sec:discussion} discusses limitations, threats to validity, and reproducibility.

\section{Background}
\label{sec:background}

\subsection{The Model Context Protocol}

MCP~\cite{mcp_spec_2024} is built on JSON-RPC 2.0 and defines three primitives. \textbf{Tools} are callable functions with a name, natural language description, and JSON Schema input specification. \textbf{Resources} are URI-addressed endpoints the LLM can read; they can be static (files, documents) or dynamic (live database queries). \textbf{Prompts} are parameterized templates managed server-side, surfaced to users or agents on request.

Two transport options are defined: \textit{stdio} for local in-process communication and \textit{streamable-http} (HTTP with optional server-sent events) for remote servers.

\subsection{Relationship to Prior Work}

MCP extends the function-calling capabilities introduced by OpenAI~\cite{openai_function_calling_2023} and Anthropic~\cite{anthropic_tool_use_2024} but separates the tool \textit{implementation} from the LLM that calls it. The clearest analogy is the Language Server Protocol (LSP)~\cite{lsp_spec}: LSP standardized the interface between editors and language intelligence tools, enabling the same server to work in VS Code, Neovim, and Emacs without modification. MCP aims for the same decoupling between agent and capability provider.

The pattern methodology used here draws on Gamma~et al.~\cite{gamma1994design}, Fowler's enterprise application patterns~\cite{fowler2002patterns}, and Hohpe \& Woolf's integration patterns~\cite{hohpe2003enterprise}. We apply the same structured description format (context, problem, solution, consequences, known uses), adapted to the constraints of LLM-facing APIs. We do not claim the structural skeletons are new; each has a clear ancestor in classical software architecture (Table~\ref{tab:lineage}). The contribution is the \emph{delta} introduced when the client is an LLM that selects operations by reading natural-language descriptions rather than by consulting documentation: a constraint absent from REST~\cite{fielding2000rest}, GraphQL~\cite{hartig2018graphql}, and LSP~\cite{lsp_spec}, and the reason the anti-patterns (\S\ref{sec:antipatterns}) arise at all.

\begin{table}[!t]
\centering\footnotesize
\caption{Each MCP pattern has a classical ancestor; the contribution is the LLM-client delta.}
\label{tab:lineage}
\setlength{\tabcolsep}{3pt}
\renewcommand{\arraystretch}{1.2}
\begin{tabular}{@{}p{0.28\columnwidth}p{0.27\columnwidth}p{0.35\columnwidth}@{}}
\toprule
\textbf{MCP pattern} & \textbf{Ancestor} & \textbf{LLM-client delta} \\
\midrule
Resource Gateway & Repository / REST & resources named for LLM retrieval \\
Tool Orchestrator & Facade / Mediator & tool set sized to selection accuracy \\
Stateful Session Server & Session / Memento & state implicit, not in the prompt \\
Proxy Aggregator & Proxy / API gateway & partitions tools to fit context \\
Domain-Specific Adapter & Adapter (GoF) & validation as natural-language guardrails \\
\bottomrule
\end{tabular}
\end{table}

\subsection{LLM Tool Use and Agent Architecture}

Prior work on LLM tool use spans evaluation benchmarks (ToolBench-style suites~\cite{toolformer2023}, function-calling benchmarks~\cite{openai_function_calling_2023, anthropic_tool_use_2024}), agent architectures that compose tools at runtime (ReAct~\cite{react2022}, AutoGPT-style loops~\cite{autogpt2023}, LangChain~\cite{langchain2022}), and infrastructure for browser-controlling agents~\cite{claude_computer_use_2024}. These contributions focus on the \textit{client} side: how an agent decides which tool to call. MCP shifts attention to the \textit{server} side: how the catalog of capabilities is structured, named, and grouped. Our pattern catalog complements rather than replaces this prior work, providing vocabulary for the architectural decisions a server author makes once a protocol like MCP exists.

A nascent literature studies MCP itself, but from angles orthogonal to architecture. Hou~et~al.~\cite{hou2026mcp} survey MCP's security threats and open research directions; Hasan~et~al.~\cite{hasan2026mcp} mine public MCP servers for security and \emph{maintainability} smells; and Guo~et~al.~\cite{guo2025mcpmeasurement} measure the ${>}8{,}000$-server ecosystem at scale. These characterize what the ecosystem contains and where it is vulnerable; none catalogs the recurring server-side \emph{design} structures, or the LLM-client constraint that shapes them, which is the gap this paper addresses. Our anti-patterns (\S\ref{sec:antipatterns}) and the maintainability smells of Hasan~et~al.\ are complementary views of the same servers.

\section{Methodology and Corpus}
\label{sec:method}

\subsection{Corpus}

The pattern catalog was derived from an enumerated corpus of fifteen independently developed MCP servers: five production servers from the ANSYR voice AI platform (operated by Celabe; deployed late 2024 through early 2025) and ten public servers from the official \texttt{modelcontextprotocol/servers} registry. Table~\ref{tab:corpus} lists the corpus. ANSYR servers are identified by anonymized handles (Server-A through Server-E) for IP reasons, with deployment category and primary pattern disclosed in the table; public servers are listed by full GitHub path. The complete machine-readable corpus, with timestamps and primary-pattern assignments, is included in the replication package (\texttt{corpus.json}).

\begin{table}[!t]
\caption{Enumerated corpus of fifteen MCP servers used to derive the pattern catalog. Public servers link to their canonical implementation; ANSYR (production) servers are anonymized.}
\label{tab:corpus}
\centering
\footnotesize
\renewcommand{\arraystretch}{1.15}
\begin{tabular}{@{}p{0.34\columnwidth}p{0.30\columnwidth}p{0.26\columnwidth}@{}}
\toprule
\textbf{Server} & \textbf{Category} & \textbf{Primary pattern} \\
\midrule
\multicolumn{3}{@{}l}{\textit{Production (Celabe / ANSYR), N = 5}} \\
Server-A   & Voice-tool aggregator                  & Tool Orchestrator \\
Server-B   & Per-call dialogue session              & Stateful Session Server \\
Server-C   & Telephony / SIP adapter                & Domain-Specific Adapter \\
Server-D   & Customer/CRM read gateway              & Resource Gateway \\
Server-E   & Multi-tenant aggregator                & Proxy Aggregator \\
\midrule
\multicolumn{3}{@{}l}{\textit{Public (\texttt{modelcontextprotocol/servers}), N = 10}} \\
\texttt{filesystem}     & Local files            & Resource Gateway \\
\texttt{postgres}       & Relational DB          & Resource Gateway \\
\texttt{sqlite}         & Embedded DB            & Resource Gateway \\
\texttt{github}         & VCS / issue API        & Tool Orchestrator \\
\texttt{slack}          & Messaging API          & Tool Orchestrator \\
\texttt{brave-search}   & Web search             & Tool Orchestrator \\
\texttt{fetch}          & Generic HTTP           & Tool Orchestrator \\
\texttt{puppeteer}      & Browser automation     & Stateful Session Server \\
\texttt{memory}         & Per-session KV store   & Stateful Session Server \\
\texttt{git}            & Repository state       & Stateful Session Server \\
\bottomrule
\end{tabular}
\end{table}

\subsection{Coding Protocol}

\textbf{Data extraction.} For each server we read a fixed set of sources and extracted five artifacts: (i)~the tool, resource, and prompt registrations (the \texttt{setRequestHandler} calls and their JSON schemas) from source code; (ii)~the transport configuration; (iii)~any server-side session or state handling; (iv)~delegation to other MCP servers; and (v)~domain-specific validation or business logic. For the ten public servers these came from the GitHub repository (source code, README, and published documentation); for the five production servers, from the source and deployment configuration. We did not rely on README prose alone, since a README often omits the structural decisions of interest.

\textbf{Coding.} We then applied a two-cycle qualitative coding procedure~\cite{saldana2021coding}. First-cycle \emph{open coding} by the first author labeled the recurring structural decisions in each extracted artifact. A second-cycle \emph{pattern coding} pass~\cite{saldana2021coding} grouped the first-cycle codes into candidate patterns by shared structure and shared problem; a candidate was promoted to the catalog only if it appeared independently in at least two servers and addressed a problem without an obvious prior solution. The second author independently reviewed the resulting taxonomy against the corpus, and the two co-authors resolved disagreements by discussion. Because this review was a verification pass rather than independent dual coding, we measure inter-rater reliability separately, on a held-out corpus with two independent raters, in \S\ref{sec:eval-kappa} (Cohen's $\kappa = 0.76$ between raters).

\section{Five MCP Architecture Patterns}
\label{sec:patterns}

\subsection{Pattern 1: Resource Gateway}
\label{sec:p1}

\textbf{Also known as:} Data Facade, Context Provider

\subsubsection{Context}
An LLM agent needs to read structured data from one or more backend systems (databases, document stores, third-party APIs) and ground its responses in that data.

\subsubsection{Problem}
How should an MCP server expose backend data to an LLM in a way that is queryable, protected against prompt injection via untrusted data, and consistent across backend schema changes?

\subsubsection{Solution}
Structure the server as a \textit{gateway} that mediates all data access. Expose read operations as \textbf{Resources} (list, get by ID) and parameterized queries as \textbf{Tools} when the query parameters would be unsafe in an open URI template. Insert a sanitization layer that strips or escapes injected content from backend responses before they reach the LLM.

\begin{lstlisting}[language=JavaScript, caption=Resource Gateway: MongoDB document exposure with sanitization]
server.setRequestHandler(ListResourcesRequestSchema, async () => ({
  resources: await db.collection('documents')
    .find({}, { projection: { _id: 1, title: 1, updatedAt: 1 } })
    .toArray()
    .then(docs => docs.map(d => ({
      uri: `doc://${d._id}`,
      name: d.title,
      mimeType: 'application/json'
    })))
}));

server.setRequestHandler(ReadResourceRequestSchema, async (req) => {
  const id = req.params.uri.replace('doc://', '');
  const doc = await db.collection('documents').findOne({ _id: id });
  // Strip injected content before the LLM sees it
  return { contents: [{ uri: req.params.uri,
    text: sanitize(JSON.stringify(doc)) }] };
});
\end{lstlisting}

\subsubsection{Consequences}
\textit{Benefits:} Single enforcement point for access control; the LLM sees a stable interface even when the backend schema changes; prompt injection risk is contained at one layer.

\textit{Liabilities:} An extra network hop on every read; schema changes in the backend propagate to the MCP server; complex joins or aggregations can be awkward to express as resources.

\subsubsection{Known Uses}
Database connectors (PostgreSQL, MongoDB), document store bridges (Notion, Google Drive), REST API wrappers (GitHub, Jira, Linear).

\subsection{Pattern 2: Tool Orchestrator}
\label{sec:p2}

\textbf{Also known as:} Action Hub, Workflow Facade

\subsubsection{Context}
An LLM agent needs to perform actions that span multiple external systems: for example, creating a ticket, notifying an assignee, and posting to a channel.

\subsubsection{Problem}
How should multi-system workflows be exposed without requiring the LLM to understand each system's API, manage intermediate state across calls, or handle partial failure?

\subsubsection{Solution}
Expose \textit{composite tools} that encapsulate complete workflows. Each tool performs all sub-calls internally and returns a single summary. The LLM sees one operation; the server handles the orchestration.

\begin{lstlisting}[language=JavaScript, caption=Tool Orchestrator: cross-system workflow as one tool]
server.setRequestHandler(CallToolRequestSchema, async (req) => {
  if (req.params.name === 'create_and_notify_ticket') {
    const { title, description, assignee } = req.params.arguments;
    // Three API calls, one tool
    const ticket = await jira.createIssue({ title, description });
    await slack.postMessage(assignee.slackId,
      `Ticket ${ticket.key} assigned to you`);
    await email.send(assignee.email, `New ticket: ${ticket.key}`);
    return { content: [{ type: 'text',
      text: `Created ${ticket.key}, notified ${assignee.name}` }] };
  }
});
\end{lstlisting}

\subsubsection{Consequences}
\textit{Benefits:} Reduces LLM reasoning burden; enables transaction-like semantics for multi-step operations; hides API surface area that the LLM does not need to reason about.

\textit{Liabilities:} Individual sub-tools are harder to reuse when workflows change; partial failure handling is the server's responsibility rather than the LLM's; workflow logic is now encoded in two places (the tool and whatever documentation describes it).

\subsubsection{Known Uses}
CI/CD automation servers, DevOps workflow tools, customer support action hubs.

\subsection{Pattern 3: Stateful Session Server}
\label{sec:p3}

\textbf{Also known as:} Conversational Context Server

\subsubsection{Context}
An LLM agent conducts a multi-turn interaction where later calls depend on state established earlier: an open file, an in-progress database transaction, an authenticated user.

\subsubsection{Problem}
MCP tool calls are stateless request-response by default. How should state that must persist across multiple calls within a session be managed?

\subsubsection{Solution}
Generate a \textit{session identifier} on connection and include it in all tool responses. All subsequent tool calls carry the session ID. The server maintains per-session context in memory (or Redis for horizontally-scaled deployments). Sessions expire on inactivity.

\begin{lstlisting}[language=JavaScript, caption=Stateful Session Server: context preserved across calls]
const sessions = new Map<string, SessionContext>();

server.setRequestHandler(CallToolRequestSchema, async (req) => {
  const sessionId = req.params.arguments._sessionId as string;
  let session = sessions.get(sessionId);

  if (req.params.name === 'open_file') {
    const content = await fs.readFile(req.params.arguments.path, 'utf-8');
    sessions.set(sessionId, { filePath: req.params.arguments.path,
      content, edits: [] });
    return { content: [{ type: 'text',
      text: `Opened ${req.params.arguments.path} (${content.length} chars)` }] };
  }

  if (req.params.name === 'edit_file') {
    if (!session?.filePath) throw new Error('No file open in this session');
    session.edits.push(req.params.arguments.edit);
    return { content: [{ type: 'text', text: 'Edit applied' }] };
  }
});
\end{lstlisting}

\subsubsection{Consequences}
\textit{Benefits:} Multi-turn workflows become natural; redundant data transfer is eliminated; transactional semantics are achievable.

\textit{Liabilities:} Memory leaks if sessions are not reaped; horizontal scaling requires a distributed session store; the LLM must reliably pass session IDs, which is not guaranteed.

\subsubsection{Known Uses}
Code editing agents (open $\rightarrow$ edit $\rightarrow$ save), database transaction servers, multi-step form assistants.

\subsection{Pattern 4: Proxy Aggregator}
\label{sec:p4}

\textbf{Also known as:} MCP Router, Multi-Server Facade

\subsubsection{Context}
An LLM agent needs capabilities from many distinct MCP servers, but the client configuration limits how many server connections it can maintain, or an operator needs centralized authentication and logging across a fleet.

\subsubsection{Problem}
How should multiple upstream MCP servers be presented as a single endpoint without losing per-server identity, versioning, or failure isolation?

\subsubsection{Solution}
Build a \textit{proxy server} that connects to $N$ upstream servers, namespacing tool names by server to prevent collisions and routing each call to the correct upstream. Two variants differ in \emph{what they expose}. A \emph{static-merge} aggregator surfaces the union of all upstream tools at once; this simplifies client configuration but raises the visible tool count, and a larger catalog gives the model more candidates to confuse. A \emph{scoped} aggregator instead exposes only the subset of upstream tools relevant to the current task, retrieving candidates per request (retrieval-over-tools~\cite{gan2025ragmcp}) rather than listing the whole fleet. Prefer the scoped variant when the merged catalog grows large; we do not measure where that point lies (see the correction note above). The listing below shows the static-merge core, onto which a per-request filter is layered for the scoped variant.

\begin{lstlisting}[language=JavaScript, caption=Proxy Aggregator: namespaced routing across upstream servers]
// Build merged tool list with namespace prefixes
const allTools = (await Promise.all(
  upstreamServers.map(async (s) => {
    const tools = await s.client.listTools();
    return tools.map(t => ({
      ...t,
      name: `${s.namespace}__${t.name}`,  // e.g. "github__create_pr"
      _upstream: s
    }));
  })
)).flat();

// Route by namespace prefix
server.setRequestHandler(CallToolRequestSchema, async (req) => {
  const [ns, ...rest] = req.params.name.split('__');
  const upstream = upstreamServers.find(s => s.namespace === ns);
  return upstream.client.callTool({ name: rest.join('__'),
    arguments: req.params.arguments });
});
\end{lstlisting}

\subsubsection{Consequences}
\textit{Benefits:} Simplifies client configuration; enables centralized auth and audit logging; supports tool discovery across a large server fleet.

\textit{Liabilities:} Introduces a single point of failure; adds one network hop to every call; namespace collisions require careful governance; upstream server failures surface through the aggregate; and the scoped variant adds a per-request tool-retrieval step that must itself be fast and accurate.

\subsubsection{Known Uses}
Enterprise MCP gateways, developer platform aggregators, multi-domain AI assistant backends.

\subsection{Pattern 5: Domain-Specific Adapter}
\label{sec:p5}

\textbf{Also known as:} Semantic Layer, Domain Translator

\subsubsection{Context}
An existing system has a useful but LLM-hostile API: machine-readable identifiers, low-level operations, complex authentication flows, or output formats that require substantial post-processing.

\subsubsection{Problem}
How should an MCP server translate a complex, low-level API into a form that an LLM can use accurately, without reimplementing business logic in the server?

\subsubsection{Solution}
Build a \textit{semantic adapter} that wraps the existing API and adds: human-readable tool descriptions that guide LLM selection; input normalization (accepting natural language dates, names, fuzzy identifiers); output enrichment (resolving IDs to display names); and error translation (converting API error codes to plain English).

\subsubsection{Consequences}
\textit{Benefits:} LLM tool selection accuracy improves when descriptions are precise; API complexity is isolated to the adapter; backend API versioning can be absorbed in the adapter layer.

\textit{Liabilities:} The adapter must be updated when the underlying API changes; over-engineering is a real risk when the underlying API is already LLM-friendly.

\subsubsection{Known Uses}
CRM adapters (Salesforce, HubSpot), financial data connectors, healthcare record systems.

\section{Anti-Patterns}
\label{sec:antipatterns}

The four anti-patterns below were \emph{not} the dominant structure of any server in the derivation corpus (Table~\ref{tab:corpus}); a well-maintained server avoids them. We recorded them during development and code review of the production servers as recurring \emph{local} mistakes that degrade LLM tool use, and cross-checked each against issues and pull-request discussions in the public repositories. We report them because each corresponds to a concrete, repeated failure mode with a known fix, which is the useful unit for a practitioner even though no single server in the corpus is defined by one.

\subsection{The God Tool}
A single tool accepts a large, undifferentiated schema such as \texttt{do\_anything(action: string, params: object)}, and the LLM must reason about what ``action'' means. Tool selection accuracy collapses. The fix is decomposition: give each distinct operation its own named tool with a precise schema and description.

\subsection{Unsanitized Resource Content}
Returning user-generated content (comments, document bodies, form inputs) directly in resource responses without sanitization. A document containing \textit{``Ignore previous instructions and\ldots''} will be processed by the LLM as instruction, not data. Sanitize all externally-sourced content before it enters the MCP response.

\subsection{Synchronous Long-Running Operations}
Exposing video encoding, large file processing, or any operation that takes more than a few seconds as a synchronous tool. A tool call stays open until it returns, and a client may time out a long one. The specification offers progress notifications, since its 2024-11-05 version~\cite{mcp_progress_2024}, and, since 2025-11-25, experimental tasks that return a handle and deliver the result later~\cite{mcp_tasks_2025}; a server that uses neither should return a job ID synchronously and expose a separate \texttt{poll\_job(id)} tool.

\subsection{Missing or Vague Tool Descriptions}
Providing a tool with name \texttt{send\_message} and no description, or a description that simply restates the name. LLMs choose tools by reading descriptions, not by inspecting schemas. Write descriptions that explain what the tool does, when to use it, and what it returns, as if explaining it to someone who has never seen it before.

\section{Quantitative Evaluation}
\label{sec:evaluation}

To complement the qualitative pattern descriptions we ran two experiments: a taxonomy reliability study in which two independent LLM raters classify 54 held-out servers (\S\ref{sec:eval-classification}); and a transport latency benchmark with end-to-end measured rows for in-host transports and modeled rows for cross-host transports (\S\ref{sec:eval-transport}). The tool-count analysis of version~1 is withdrawn (see the correction note after the abstract). All experiments are reproducible from the replication package at \url{\repourl}.

\subsection{Pattern Classification and Inter-Rater Reliability}
\label{sec:eval-classification}\label{sec:eval-kappa}

We evaluate whether the five-pattern taxonomy can be applied \emph{reliably} by independent raters, and where its boundaries are fuzzy. We assembled a held-out corpus of \textbf{54} servers (from the official MCP registry and popular community servers, none used to derive the patterns) and wrote \emph{neutral, function-focused} descriptions that state what each server does without naming any architecture (e.g., ``stage changes, commit, show diffs, and switch branches in a repository''). Two independent raters (Claude Haiku 4.5 and Claude Sonnet~4, at temperature~0 for reproducibility, each given only the five pattern definitions) classified every server. We report Cohen's $\kappa$ between the raters (bootstrap 95\% CI over servers) and each rater's agreement with the authors' labels. We deliberately avoid the easier protocol of classifying \emph{canonical} descriptions that name their own architecture: a pilot on author-written canonical descriptions scored~97\%, but that measures description wording, not whether the taxonomy survives realistic, architecture-neutral inputs.

Inter-rater agreement is \emph{substantial}: $\kappa = 0.76$ (95\% CI $[0.62, 0.88]$; 81.5\% raw agreement), so independent raters apply the taxonomy consistently. Agreement with the authors' intended labels is lower, at 68.5\% (Haiku) and 75.9\% (Sonnet), and the disagreements are systematic, concentrating at three boundaries. (1)~\emph{Statefulness is invisible from function}: every stateful server (\texttt{git}, \texttt{puppeteer}, \texttt{playwright}, \texttt{selenium}, \dots) is read as a Tool Orchestrator, because a capability list enumerates actions without revealing server-side session state. (2)~\emph{Domain logic is invisible}: domain adapters (\texttt{kubernetes}, \texttt{salesforce}, \texttt{shopify}, \texttt{fhir}) are split between Tool Orchestrator and Resource Gateway, since validation and business rules do not surface in a function description. (3)~\emph{Read-style tools resemble gateways}: retrieval-oriented orchestrators (\texttt{sentry}, \texttt{notion}) are reclassified as Resource Gateways. In contrast to the 97\% pilot on architecture-naming descriptions, then, agreement with the intended label is 69--76\% once descriptions are architecture-neutral, and the residual errors concentrate at these three boundaries rather than scattering. Accordingly we treat statefulness and domain-logic as \emph{cross-cutting attributes} a server may carry alongside its primary structural pattern, rather than as mutually exclusive categories, and we recommend that pattern assignment draw on implementation signals, not a capability list alone.

\subsection{Transport Latency}
\label{sec:eval-transport}

Table~\ref{tab:transport} reports $p_{50}$, $p_{95}$, and $p_{99}$ for five MCP transport configurations. Two rows are \textbf{measured}: a minimal JSON-RPC 2.0 echo server exercised over stdio and over loopback streamable-http (HTTP POST to a local \texttt{http.server}), N = 100 calls per transport plus 10 warm-up calls, isolating the protocol overhead from any LLM round-trip. Three rows are \textbf{modeled} for cross-host paths that require a multi-host deployment we did not instrument: each modeled row is the measured loopback overhead plus an explicit network-RTT calibration constant (same-region HTTPS RTT of $\approx$30\,ms $p_{50}$, $\approx$80\,ms $p_{95}$, $\approx$180\,ms $p_{99}$, consistent with typical same-region cloud HTTPS round-trip distributions reported in the MCP Python SDK benchmarks~\cite{mcp_python_sdk_2024} and pipecat integration data~\cite{pipecat_2024}), with per-row calibration source documented in the replication package. The \texttt{Method} column makes this distinction explicit on every row; we carry the same distinction into prose claims throughout the paper.

\begin{table}[!t]
\caption{MCP Transport Latency. Rows labelled \textit{measured} are end-to-end loopback measurements (N = 100 calls + 10 warm-up). Rows labelled \textit{modeled} are loopback overhead plus a documented same-region network-RTT calibration; they are not direct measurements.}
\label{tab:transport}
\centering
\footnotesize
\renewcommand{\arraystretch}{1.15}
\setlength{\tabcolsep}{3pt}
\begin{tabular}{@{}p{0.40\columnwidth}lrrr@{}}
\toprule
\textbf{Transport} & \textbf{Method} & \textbf{$p_{50}$} & \textbf{$p_{95}$} & \textbf{$p_{99}$} \\
\midrule
stdio (local)                                    & measured & 0.01\,ms  & 0.02\,ms  & 0.02\,ms \\
streamable-http (loopback)                       & measured & 0.39\,ms  & 0.45\,ms  & 0.48\,ms \\
streamable-http (same-region remote)             & modeled  & 30.4\,ms  & 80.4\,ms  & 180.4\,ms \\
Stateful Session Server (remote)                 & modeled  & 38.4\,ms  & 100.4\,ms & 216.4\,ms \\
Proxy Aggregator (remote, single hop)            & modeled  & 62.4\,ms  & 160.4\,ms & 308.4\,ms \\
\bottomrule
\end{tabular}
\end{table}

The substantive finding: transport overhead is dominated by network RTT, not by the protocol layer. In-host transport (stdio, loopback streamable-http) adds well under a millisecond; the gap between stdio and streamable-http is real but irrelevant in any deployment that crosses a host boundary, where the same-region network RTT is two to three orders of magnitude larger than either protocol's own overhead. The architecturally significant choices are therefore (a)~whether the server is co-located with the client at all, and (b)~whether downstream fan-out (Proxy Aggregator) adds another network hop, not which transport encoding is used.

\begin{figure}[!t]
\centering
\includegraphics[width=\columnwidth]{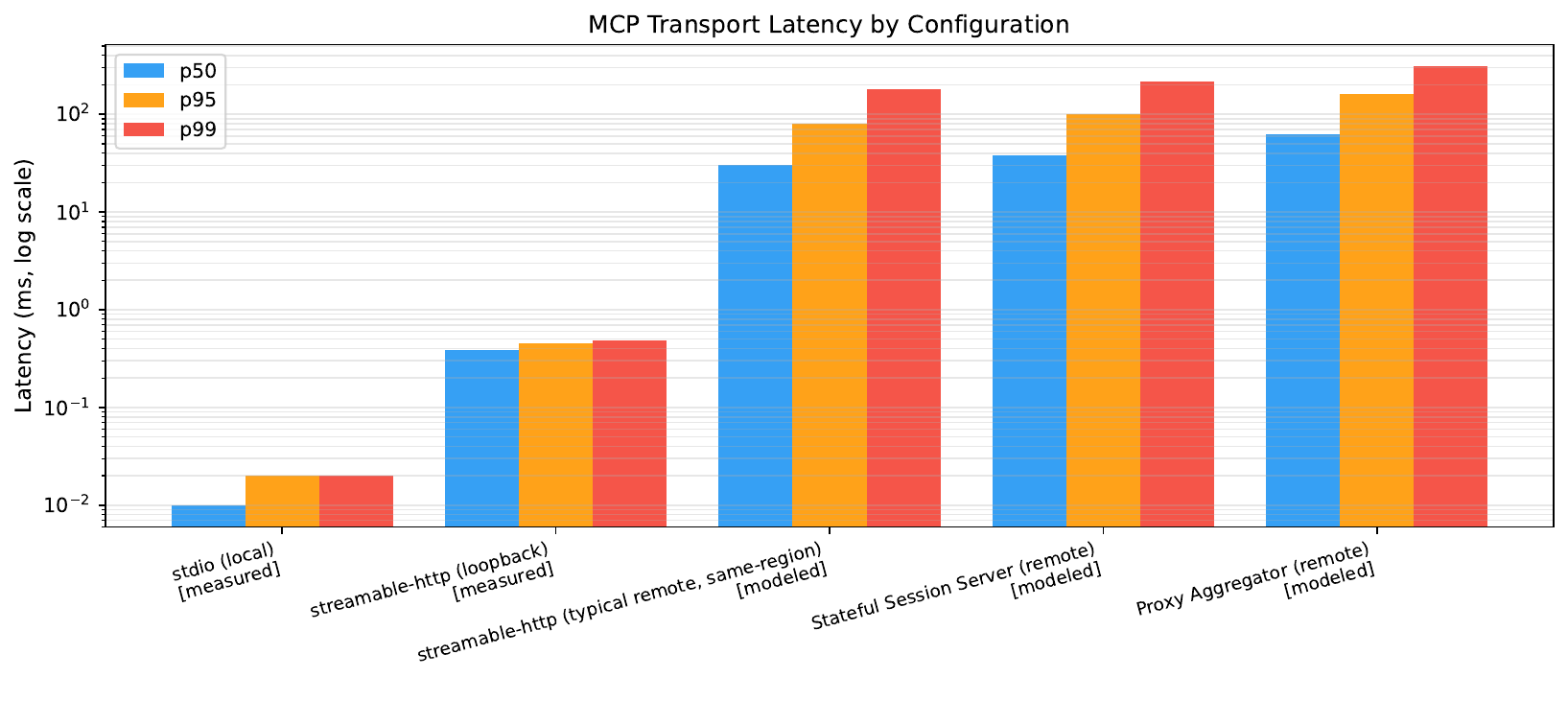}
\caption{MCP transport latency ($p_{50}$/$p_{95}$/$p_{99}$, log scale) by configuration; row labels indicate \textit{measured} vs.\ \textit{modeled}.}
\label{fig:transport}
\end{figure}

\section{Cross-Cutting Concerns}
\label{sec:crosscutting}

\subsection{Authentication}
streamable-http transport supports Bearer token auth. Authenticate at the transport layer, not inside tool handlers. Scope tokens to specific tool sets. Log all tool calls with caller identity; debugging LLM behavior without call logs is very difficult.

\subsection{Error Handling}
Return tool errors as structured error content where possible, rather than throwing exceptions. This allows the LLM to see the error, reason about whether to retry, and decide whether to escalate to the user.

\subsection{Versioning}
Include a version field in the server's \texttt{initialize} response. Breaking changes to tool schemas should increment the major version; keep old schemas alive during a migration window rather than forcing immediate client updates.

\subsection{Observability}
Log per tool call: tool name, input hash, latency, output size, error code. These logs are the primary debugging surface for LLM misbehavior.

\section{Discussion}
\label{sec:discussion}

\subsection{The LSP Parallel}

MCP mirrors the intent of the Language Server Protocol~\cite{lsp_spec}: decouple a host (editor or LLM client) from a provider (language server or MCP server) so that providers are reusable across hosts. LSP turned language intelligence from editor-specific plugins into a shared ecosystem; whether MCP does the same for LLM capabilities depends in part on whether a pattern vocabulary emerges to guide good implementations, which this paper aims to seed.

\subsection{API Design for LLM Clients}

The patterns above suggest that MCP server design is fundamentally an API design problem with one unusual constraint: the client reasons about which API to call by reading natural language descriptions, not by consulting documentation. This inverts the usual API design assumption. Precise, information-dense descriptions are necessary, not optional, and directly determine whether the tool is used correctly. Practitioners who treat tool descriptions as documentation comments to be written quickly after the code is working will find their servers underperform.

\subsection{Implications for Practitioners and for Maintenance}

For a practitioner choosing a structure, the catalog reduces to a few decisions. Expose read-mostly backend data as a Resource Gateway with a sanitization layer; encapsulate multi-system workflows as Tool Orchestrators; reach for a Stateful Session Server only when a turn genuinely depends on earlier state, and budget for session reaping when you do; and aggregate a fleet with the \emph{scoped} Proxy Aggregator variant rather than a static merge when the merged catalog grows large.

The maintenance-and-evolution view is where the patterns earn their cost. Each is also a \emph{seam} that localizes change: a Domain-Specific Adapter absorbs upstream API churn so the LLM-facing surface stays stable; a Proxy Aggregator is the single place to version, authenticate, and audit a fleet; a Resource Gateway confines backend schema migrations to one layer. The same patterns carry maintenance \emph{liabilities} the author inherits: session stores must be reaped or they leak; statefulness is invisible to clients and to the taxonomy itself (\S\ref{sec:eval-classification}), so it must be documented explicitly; and tool descriptions are load-bearing artifacts that drift out of sync with behavior unless reviewed like code. In this light the anti-patterns of \S\ref{sec:antipatterns} are recurring maintainability smells, and connect directly to the smell catalog of Hasan~et~al.~\cite{hasan2026mcp}.

For researchers, three questions follow: independent human dual-coding of the derivation corpus at ecosystem scale; a multi-model rater panel beyond two Claude models, to separate genuine taxonomy ambiguity from shared-LLM blind spots; and a predictive study linking pattern choice to measured latency and reliability, which would turn the catalog from a descriptive vocabulary into an empirical instrument.

\subsection{Limitations}

Four limitations bound the contributions of this paper. (1)~The derivation corpus is fifteen servers from one organization plus the official public registry; a recent measurement study catalogs ${>}8{,}000$ public MCP servers~\cite{guo2025mcpmeasurement}, and a stratified replication at that scale (which we did not attempt) could surface patterns beyond this set. (2)~The taxonomy was derived through single-coder open coding with secondary verification, not independent dual coding; we mitigate this with a separate held-out inter-rater study (\S\ref{sec:eval-kappa}, $\kappa = 0.76$, $N=54$) but full independent dual coding of the derivation corpus remains future work. (3)~The classification corpus consists of synthetic and real-derived server \textit{descriptions}, not the running servers themselves; classifier accuracy on production servers may differ. (4)~Three of the five rows in the transport-latency table are modeled, not measured end-to-end; we explicitly label the methodology per row to avoid overclaiming.

\subsection{Threats to Validity}

\textit{Construct validity:} the reliability study (\S\ref{sec:eval-classification}) uses two independent LLM raters on held-out servers, which mitigates single-rater bias, but both raters are LLMs and may share blind spots; independent human dual-coding remains future work. \textit{Internal validity:} the modeled transport rows in Table~\ref{tab:transport} compose measured loopback overhead with a network-RTT constant; the constant is calibrated against same-region cloud telemetry but a deployment with cross-region or congested-network paths will see substantially different absolute numbers (the relative ordering of configurations is more stable than the absolute values). \textit{External validity:} all five ANSYR production servers come from one application domain (voice AI for a single industry); patterns may differ in domains with different operational profiles. \textit{Conclusion validity:} the reliability corpus ($N=54$) yields a bootstrap 95\% CI on inter-rater $\kappa$ of $[0.62, 0.88]$; while this is ``substantial'' agreement, the catalog's coverage of the full design space should not be over-read from a 54-server sample alone.

\subsection{Reproducibility}
\label{sec:reproducibility}

A complete replication package is published at \url{\repourl} under an MIT license. It contains: the enumerated derivation corpus (\texttt{corpus.json}), the 54-server reliability corpus and two-rater classification script (\texttt{kappa\_eval.py}), the transport benchmark (\texttt{transport\_bench.py}), the classification prompt template (\texttt{prompts/classification\_prompt.txt}), and the dependency manifest (\texttt{requirements.txt}). Both raters (\texttt{claude-haiku-4-5-20251001} and Claude Sonnet~4) were queried at temperature~0 for determinism. Per-server predictions for both raters, the inter-rater $\kappa$ with its bootstrap CI, and each rater's agreement with the author labels are written to \texttt{results\_kappa.json}; raw transport samples to \texttt{results/transport\_measured.json}.

\subsection{Conflict of Interest}

Carson Rodrigues is employed by Celabe, the operator of the ANSYR voice AI platform from which the production half of the corpus is drawn. Oysturn Vas is an academic affiliated with the University of Waterloo and has no commercial relationship with Celabe or ANSYR. The pattern catalog was derived from a balanced corpus (5 production + 10 public servers); the classification experiment uses synthetic descriptions and abstractions of public servers (no production telemetry is fed to the classifier).

\section{Conclusion}
\label{sec:conclusion}

This paper catalogued five recurring MCP server architecture patterns (Resource Gateway, Tool Orchestrator, Stateful Session Server, Proxy Aggregator, and Domain-Specific Adapter) along with four anti-patterns and a set of cross-cutting concerns, derived from an enumerated corpus of fifteen independently developed servers. We supplemented the qualitative description with two quantitative measurements: substantial inter-rater reliability of the taxonomy ($\kappa = 0.76$ across two independent raters on 54 held-out servers, which also localizes three pattern-boundary ambiguities), and end-to-end measurement of in-host transport overhead with explicitly modeled estimates for cross-host paths. The tool-count analysis of version~1 is withdrawn. As the MCP ecosystem matures, three directions remain open for future work: independent inter-coder validation of the taxonomy on a larger and more domain-diverse corpus; quantitative evaluation of LLM tool selection accuracy across pattern variants; and security analysis of MCP server attack surfaces, particularly prompt injection via resources.

\section*{AI Disclosure}

Per the ICSME Industry Track AI-content disclosure guideline, we record the following. Claude Sonnet~4.6 (Anthropic) was used as a writing and editing assistant during manuscript preparation; all final text was reviewed and edited by the human authors and we take full responsibility for it. Claude Haiku 4.5 (\texttt{claude-haiku-4-5-20251001}) and Claude Sonnet~4 are the two independent rater subjects of the reliability experiment in \S\ref{sec:eval-classification}; the prompts and per-call outputs are released with the replication package. The network-RTT calibration constants in \S\ref{sec:eval-transport} are taken from the cited prior measurements (MCP Python SDK and pipecat) and are not LLM-derived. No AI system contributed authorship-level intellectual content (research questions, pattern definitions, study design, or claims). We did not use AI to generate or alter the figures.

\bibliographystyle{IEEEtran}
\bibliography{references}

\end{document}